\documentclass[12pt,a4paper]{article} 
\usepackage[utf8]{inputenc}
\usepackage[english]{babel}
\usepackage{amsmath}
\usepackage{amsfonts}
\usepackage{amssymb}
\usepackage{graphicx}
\usepackage[font=footnotesize,labelfont=bf]{caption}
\usepackage{authblk}
\usepackage[colorlinks=true,citecolor=blue]{hyperref}

\usepackage{subcaption}

\usepackage[]{natbib} 
\usepackage{color}
\newcommand{\tr}{\color{red}}

\begin{document}
\renewcommand{\abstractname}{\vspace{-2\baselineskip}}

\title{Three-dimensional oscillatory magnetic reconnection}
\author[1,2]{J.~O.~Thurgood\textsuperscript{*}}
\author[2]{D.~I. Pontin}
\author[1]{J.~A.~McLaughlin}
\affil[1]{Department of Mathematics, Physics and Electrical Engineering, Northumbria University, UK.}
\affil[2]{Division of Mathematics, University of Dundee, UK.}
\affil[]{jonathan.thurgood@northumbria.ac.uk}
\date{}
\setcounter{Maxaffil}{0}
\renewcommand\Affilfont{\itshape\small}
\maketitle

\abstract{
    \textbf{Here we detail the dynamic evolution of localised reconnection regions about three-dimensional (3D) magnetic null points by using numerical simulation.
We demonstrate for the first time that reconnection triggered by the localised collapse of a 3D  null point due to an external MHD wave involves a self-generated oscillation, whereby the current sheet and outflow jets undergo a reconnection reversal process during which back-pressure formation at the jet heads acts to prise open the collapsed field before overshooting the equilibrium into an opposite-polarity configuration. 
The discovery that reconnection at fully 3D nulls can proceed naturally in a time-dependent and periodic fashion is suggestive that oscillatory reconnection mechanisms may play a role in explaining periodicity in astrophysical phenomena associated with magnetic reconnection, such as the observed quasi-periodicity of solar and stellar flare emission.
Furthermore, we find a consequence of oscillatory reconnection is the generation of a plethora of freely-propagating MHD	 waves which escape the vicinity of the reconnection region}
}
\vspace*{1cm}
\linebreak
{\tr{Manuscript in press, accepted for publication by ApJ in June 2017. The final published version will be available with 'gold' open access, see the main journal for access to supplementary animations.}}

\section{Introduction}\label{sec:intro}

Magnetic fields play a key role in determining the dynamics of plasmas at all scales: from fusion experiments and laboratory plasmas, to planetary magnetospheres, the Sun and stars, to galaxies and accretion disks. Magnetic reconnection is a fundamental plasma process associated with dynamic energy release in these systems, and is believed to explain a broad range of phenomena including solar and stellar flares, coronal mass ejections, astrophysical jets and planetary aurorae. Current theoretical frontiers in fundamental reconnection research can be broadly categorised as; (i) {collisionless}/kinetic effects \citep{2010RvMP...82..603Y}, (ii) the extension to three dimensions (3D) of a long history of two-dimensional (2D) theory (where profound differences make this extension highly non-trivial \citep{priest2003a,Pontin3169}), (iii) time-dependent and transient effects, and (iv) the interaction of the local reconnection dynamics with those of the global system. Here, we present significant results concerning the latter three within the framework of nonlinear magnetohydrodynamics (MHD).

{Much of} our present understanding of reconnection has focused primarily on continuously-driven (and hence at least quasi-steady) systems, whilst transient effects that are crucial in many plasma environments are typically neglected {(we note an important exception in the case of current sheet instabilities, which are outside of the scope of this paper, where studies are necessarily time-dependent, e.g. \citet{2007PhPl...14j0703L,2016PhPl...23j0702C})}. Combined with questions regarding the influence of chosen boundary conditions, the applicability of any given reconnection model to a particular situation is often unclear. Here we study instead the self-consistent formation and evolution of a reconnecting current sheet in response to a finite external perturbation. The reconnection is triggered in the vicinity of a null point, a natural \lq{weakness}\rq{ } in the magnetic field, by an external driver in the form of an MHD wave. It is well established that such null points trap these waves due to refraction \citep{2011SSRv..158..205M} leading to the formation of electric current layers around the null in which reconnection may take place {\citep{pontincraig2005,pontinbhat2007a,2009A&A...493..227M}}.
What has so far not been studied is the subsequent reconnection dynamics triggered by the waves in a configuration where the reconnection region is  connected naturally to an external, non-reconnecting region. That is the aim of this paper.

Within the reduced framework of 2D MHD, 
where the assumption of an invariant direction simplifies the system greatly,
 it has recently been demonstrated that in such a scenario reconnection does not proceed in a quasi-steady manner but rather exhibits a time-dependent behaviour, termed
\textit{Oscillatory Reconnection} \citep{1991ApJ...371L..41C,1993ApJ...417..748O,2009A&A...493..227M,2009A&A...494..329M,
2012ApJ...749...30M,2014ApJ...796...43P}.
However, the extension to {3D} is not trivial; {the process by which waves trigger reconnection (\textit{null collapse}, see e.g. Chapter 7 of \citealt{priestandforbes}) is little-studied in three-dimensions,} and {also} reconnection itself {can be} fundamentally changed \citep{priest2003a}. {Additionally,} mode conversion between magnetoacoustic and Alfv\'en waves becomes permissible. 

Here we demonstrate for the first time that {impulsively initiated} reconnection at realistic 3D magnetic null points naturally proceeds in a time-dependent and oscillatory fashion, and also demonstrate
that 3D reconnection can itself produce MHD waves.
The results show that reconnection is inherently periodic and highlight the deeply-interconnected nature of MHD waves and reconnection.

\section{Simulation Setup}\label{setup}

\begin{figure*}
    \centering
    \includegraphics[width=0.875\linewidth]{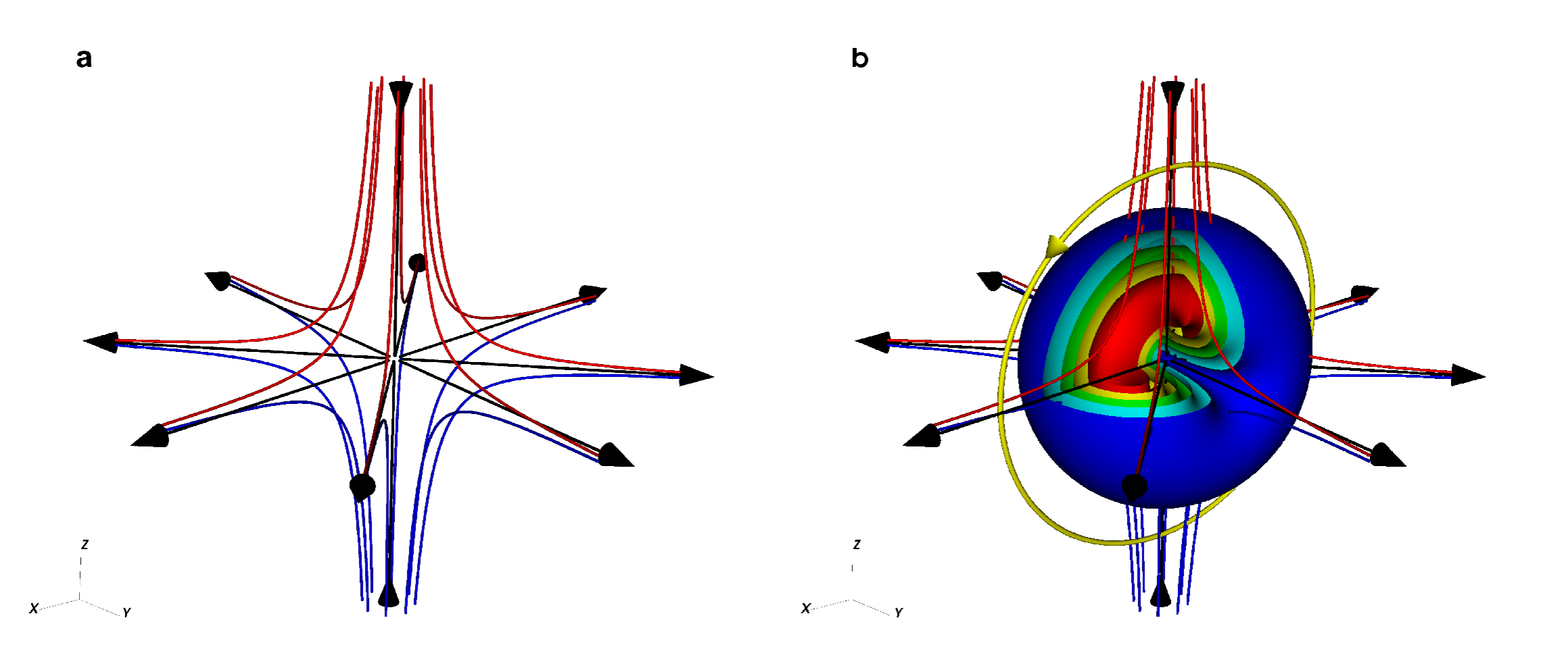} 
\caption{
  	\textbf{(a)} The  equilibrium magnetic field $\mathbf{B}_{0}$.
  	 Black fieldlines illustrate the behaviour of the spine and fan fieldlines, where the \textit{fan} fieldlines point radial from the null in the $z=0$ plane and the \textit{spine} line which points towards the null along the $z$-axis. Red fieldlines are traced from above the fan, and blue from below.
  	\textbf{(b)} The perturbing flux ring $\mathbf{B}_{1}$, superimposed upon the background field. Coloured isosurfaces profile the increasing perturbation field strength from zero (transparent) through weak (blue) to strong (red). The circulation of flux ring fieldlines in planes of fixed $y$ is illustrated by the yellow line.   
\label{fig:setup}}
\end{figure*}

The simulation involves the numerical solution of the MHD equations using the \textit{Lare3d} code \citep{2001JCoPh.171..151A}. Here we outline the initial conditions, with full technical details deferred to the appendix. All variables in this paper are nondimensionalised. We consider a background magnetic field of the form:
\begin{equation}\label{b0}
\mathbf{B}_{0} = \left[x,y,-2z \right] \,,
\end{equation}
which is known as the \textit{linear, proper potential null point} \citep{1996PhPl....3..759P}. The magnetic null point itself ($\mathbf{B}=\mathbf{0}$) is located at the origin. The field is free from electrical currents and so constitutes a minimum magnetic energy state.
Topologically, it consists of a \textit{spine} fieldline, running along the $z$-axis  towards the null point, and a \textit{fan plane} $z=0$, consisting of fieldlines pointing radially outwards. Other fieldlines, separated by the fan, have a hyperbolic structure (Figure \ref{fig:setup}a).

Upon this magnetic field we impose a finite-amplitude perturbation $\mathbf{B}_{1}=\mathbf{\nabla}\times\mathbf{A}_{1}$ where 
\begin{equation}\label{b1}
\mathbf{A}_{1} = \psi\,\mathrm{exp}\left[\frac{-\left(x^2+y^2+z^2\right)}{2\sigma^2}\right] \hat{\mathbf{y}} \,,
\end{equation}
which is a ring of magnetic flux centred about the null point, circulating in planes of fixed $y$ (Figure \ref{fig:setup}b).
When the simulation begins, the perturbation immediately splits into inward and outward propagating magnetoacoustic waves.
The incoming part is our intended driver for initiating magnetic reconnection and acts implosively to collapse the magnetic field into the first reconnection configuration.
The outgoing part of the perturbation leaves the initialisation site and propagates outwards.
Special care is taken so that it and any subsequently generated waves do not reflect at the boundaries and return inwards to influence our solution at the null ({see the appendix}).

The background plasma is taken to be initially stationary ($\mathbf{v}=\mathbf{0}$) and of uniform density ($\rho=1$).  We are now left with a choice of a value to assign to the perturbation energy (set by $\psi$ and $\sigma$), and the uniform pressure $p$ and resistivity $\eta$. As ${\bf B}_{0}$ is scale-free the relative sizes of these three parameters control the dynamics of the initial implosion.
{Here we  present results for $\psi=0.05$, $\sigma=0.21$, $\eta=10^{-3}$, and $p=0.005$. This choice of parameters is such that the implosion forms a localised current sheet which is isolated from the boundaries and free to evolve in a self-consistent manner. 
  }

\section{Results}\label{results}
\subsection*{Initial implosion}

\begin{figure*}
    \centering
    \includegraphics[width=0.875\linewidth]{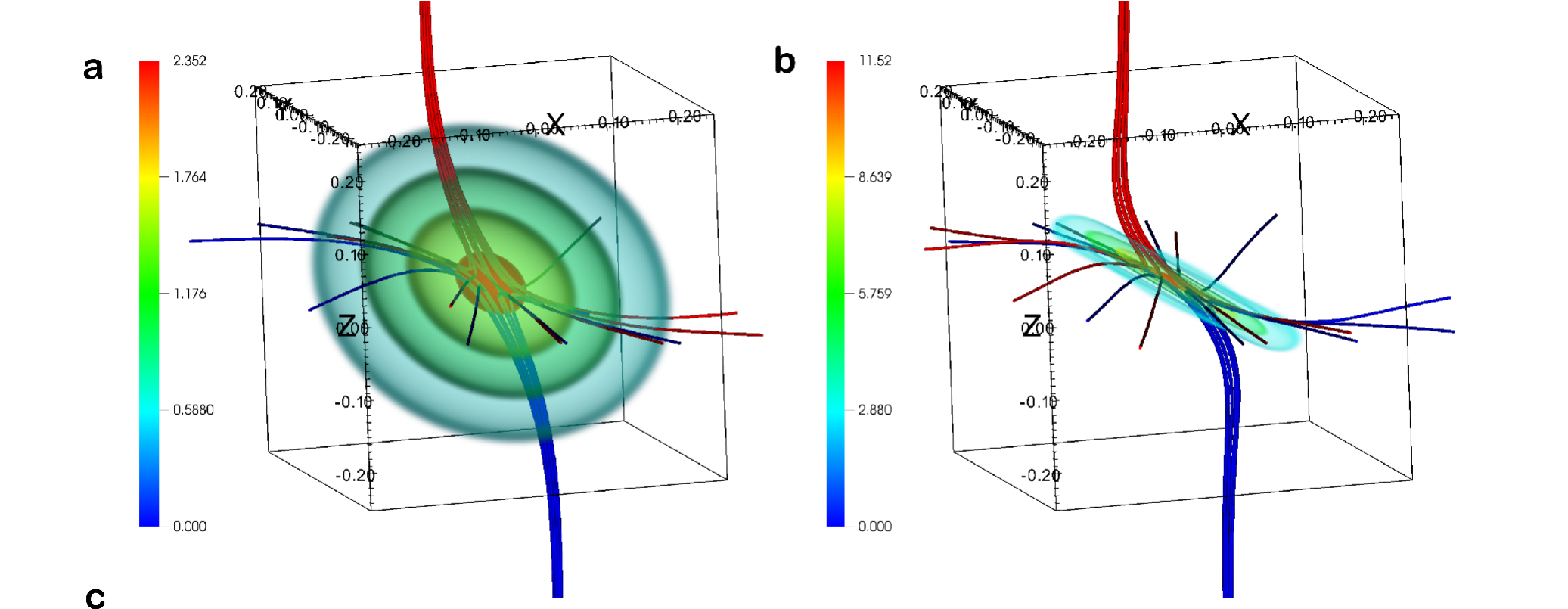} 
\caption{Transparent isosurfaces illustrating 3D distribution of $j_y$ about the collapsed field close to the null at $t=0.1$  \textbf{(a)} and \textbf{(b)} $t=0.5$. Note the changing colour scales.
} \label{fig:current}
\end{figure*}
\begin{figure*}
    \centering
    \includegraphics[width=0.875\linewidth]{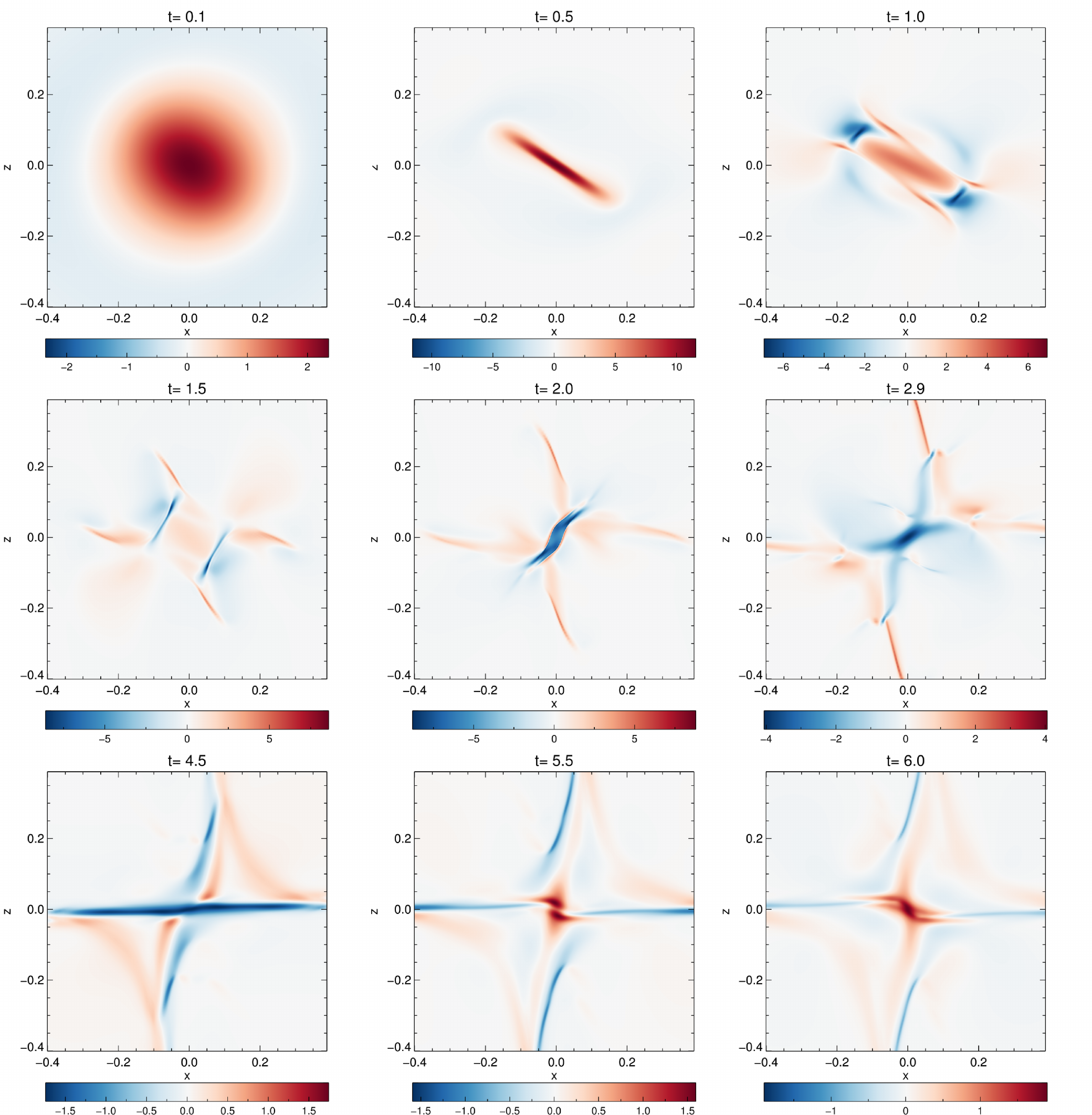} 
\caption{Longer term evolution of $j_y$ in the $y=0$ plane, showing the oscillation of the sheet. Note the changing colour scales. See also the supplementary animation.
} \label{fig:current2}
\end{figure*}

The incoming wave is initially symmetric as per the geometry of the flux ring and generates a cylindrical ring current flowing through the null. It propagates towards the null point as a fast magnetoacoustic mode. In  linearised MHD it would evolve in a self-similar manner, focusing its energy towards the null, increasing in amplitude. Due to the linearly-decreasing Alfv\'en speed profile ($c_{A}\propto|B_{0}|$) length scales decrease exponentially, and so current density at the null point increases exponentially in time {in the linear regime} whilst retaining its cylindrical symmetry \citep{2004A&A...420.1129M,2008SoPh..251..563M}. This focusing would continue until either resistive diffusion or plasma back-pressure act to halt further collapse, the scalings for which are well-understood only in 2D  \citep{1991ApJ...371L..41C,1992ApJ...399..159H,1993ApJ...405..207C,
1992ApJ...393..385C,1996ApJ...466..487M}. However, given that our simulation considers a full nonlinear solution, there exists the opportunity for increasing amplitudes to allow the excess flux carried by the wave to overwhelm the background field and so form shocks.

The  approaching pulse imbalances the distribution of magnetic flux about the null point, enhancing and diminishing the overall field strength in different lobes about the null point's spine and fan. The anisotropic nature of the Lorentz force leads  to a net force that is directed alternately towards and away from the null point, in the regions of magnetic enhancement and rarefaction respectively.  
It follows that the implosion carries fronts of magnetic and plasma compression and rarefaction about the spine and fan. 
For a sufficiently energetic perturbation (such as the one considered here) nonlinear steepening occurs in these compressive fronts, leading to a breaking of the cylindrical symmetry. Past studies of 2D null collapse demonstrated that the implosion becomes quasi-1D  as its fronts accelerate and stall in regions of compression and rarefaction, leading to collapse of the separatricies towards one another \citep{1992ApJ...393..385C,1996ApJ...466..487M,2011A&A...531A..63G}.

An analogous breaking of the symmetry is observed in our simulations, with the disturbance flattening into a planar structure. Figure \ref{fig:current} shows selected transparent isosurfaces of the $y$-component of current at an early time ($t=0.1$) and at the approximate time at which the collapse is halted ($t=0.6$). The $y$-component of current is that which is flowing through the null, generating the parallel electric field required for 3D reconnection.  We see that at $t=0.1$, the current profile has already started to depart from the initial cylindrical symmetry in planes of fixed $y$, and by $t=0.6$ a quasi-planar current sheet has been formed by the aforementioned processes of nonlinear steepening in the compressive fronts. The increase in amplitude is due to the wave focusing. Thus, the perturbation creates a current sheet which is localised in the vicinity of the null point 
{and so we find that 3D null collapse, for our choice of parameters, is qualitatively similar to 2D collapse in planes of fixed-$y$. It differs in that the current distribution is however fully localised in 3D, as can be seen in Figures \ref{fig:current}a and b. As we will see in the following section, there is a crucial difference in that the 3D collapse triggers a 3D reconnection mode, which is significantly different to 2D}. We highlight that this sheet is detached from the influence of the boundaries and has been generated in an impulsive rather than driven manner -- once the current sheet is formed, no further energy is injected to sustain the sheet or otherwise alter it.

\subsection*{Reconnection reversals}
\begin{figure*}
    \centering
    \includegraphics[width=0.875\linewidth]{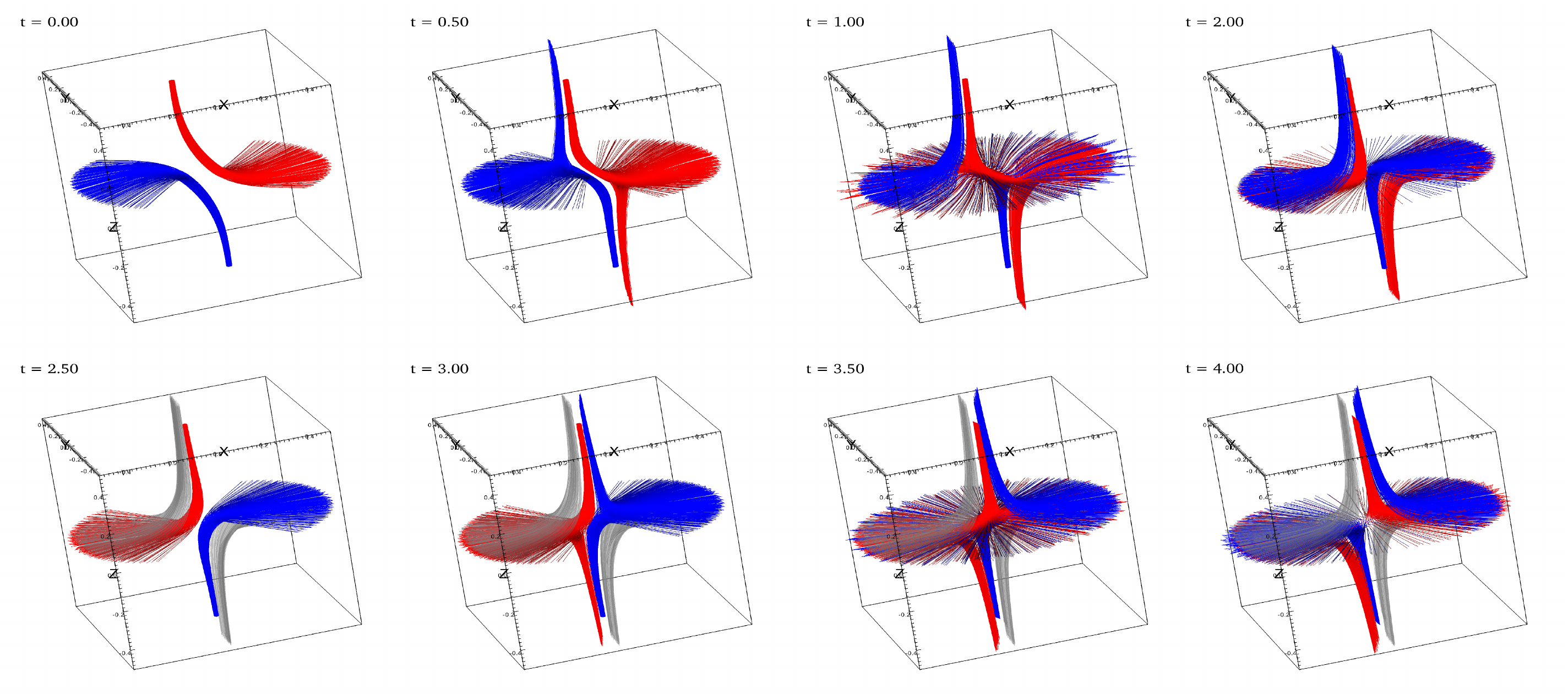} 
\caption{ The reconnection of two thin flux tubes which are initially symmetric about either side of the initial current sheet is shown. The top row spans the first reconnection event (positive current at the null), where fieldlines anchored above and below the null (i.e. at larger $z$) flip around the spine, and fieldlines anchored closer to the fan plane (i.e at lower $z$) flip up the spine, thus being transferred through the fan plane. This connectivity change is \textit{spine-fan reconnection}. By the completion of the first reversal the selected flux tubes have bifurcated into four separate tubes. The bottom row follows them through the second reversal. The exterior flux tubes do not participate in the second reconnection event and are so recoloured as a transparent grey for clarity. The interior flux ropes are again reconnected in the spine-fan mode, but in the opposite direction to the initial reconnection due to the current reversal. See also the corresponding supplementary animation.
} \label{fig:rcn}
\end{figure*}

The longer term evolution of the current flowing through the plane of collapse ($j_{y}$ in the $y=0$ plane) is shown in Figure \ref{fig:current2}. In the first few frames, we see the formation of a current sheet by $t\approx0.6$ due to the nonlinear implosion collapsing the null point's spine and fan towards each other (note the { spine has locally `collapsed' onto the fan} in Figure \ref{fig:current}b). Shortly after the initial implosion halts, strong opposite { (negative)} polarity currents form at the ends of the current sheet and begin to propagate inwards towards the null point. These current concentrations -- called \textit{deflection currents} due to their {close} association with the formation of a shock in the reconnection outflow {(see \citealt{1986ApJ...305..553F})}-- act to prise open the collapsed spine and fan fieldlines and so relieve the force imbalance about the null point which supports the pre-existing (positive) current sheet. The pre-existing current sheet thus shortens along what was its length-wise axis and broadens along its width-wise axis, with a decrease in the current density at the null point itself. Eventually, this opposite polarity current completely reverses the spine-fan collapse, and overshoots, causing a collapse of the { spine and fan magnetic fieldline structures} in the opposite direction. The { deflection current} concentrations thus eventually coalesce to form a new current sheet by around $t\approx2$. The process then repeats, restoring the sheet to its original polarity before the simulation end time of $t=6$. The dynamics of these \textit{reconnection reversals} are considered later.

We confirm magnetic reconnection occurs across these oscillating current sheets and investigate its nature by tracking the changes to fieldline connectivity within selected magnetic flux tubes during the simulation (Figure \ref{fig:rcn}). Initially, we see that the two selected fieldline bundles are carried in towards the null by the wave-generated inflow. As they enter the vicinity of the current sheet (diffusion region), the pairwise magnetic connectivity between the fluid elements at ends of the ropes is lost as fieldlines are continuously reconnected across both the spine line and fan plane. 
{In this fashion, the reconnection resulting from 3D null collapse is distinct from the 2D case. Unlike in 2D reconnection, where connectivity changes occur always in a pair-wise sense at the null point only, here connectivity is changed instant-to-instant throughout the whole of the non-ideal region \citep{priest2003a}.}
 The observed connectivity change is indicative of the \textit{spine-fan} mode of 3D reconnection  \citep{2005GApFD..99...77P,2009PhPl...16l2101P}, where magnetic flux is redistributed across both the spine and fan plane.
By the point at which the first reversal has completed ($t\approx2$, cf. Figures \ref{fig:current2} and \ref{fig:rcn}) the selected flux ropes have been fully redistributed into opposite lobes about the spine and fan. 
{As all magnetic connectivity is lost between the four groups of seed particles at the ends of the selected flux ropes during the first reconnection event (see Appendix \ref{seedsapp})}, the selected flux ropes are bifurcated into four distinct ropes of flux by {the completion of the first reversal}. 
As time further progresses during the period of negative current at the null point,  the inner-most of the selected flux ropes are reconnected through the diffusion region again, but in the opposite direction due to the reversal of current.  The figure therefore demonstrates that the transfer of flux about the spine and fan due to spine-fan reconnection is oscillatory and time-dependent.

\subsection*{Reconnection jets and reversal dynamics}
\begin{figure*}
    \centering
    \includegraphics[width=0.875\linewidth]{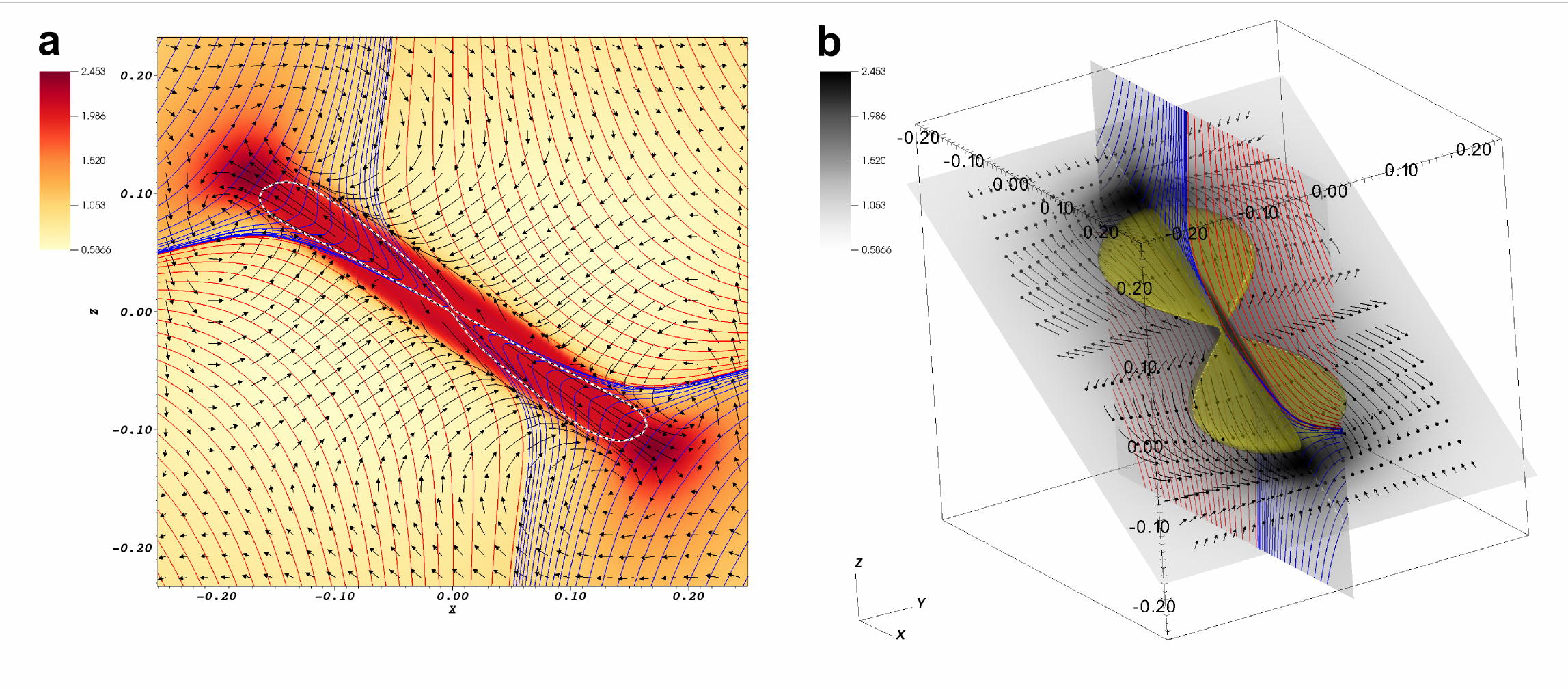} 
\caption{The reconnection jet at $t=0.7$. \textbf{(a)} The system of flow in the $y=0$ plane, where the colour map illustrates density, blue curves are magnetic fieldlines in the outflow {regions}, and red {curves are fieldlines} in the inflow {regions} ({with the inflow and outflow regions} separated by the spine and fan). Black velocity arrows indicate flow direction and magnitude. The dashed {white} contour encloses the region of super-magnetosonic flow ($M_{A}>1$).
\textbf{(b)} The system of flow out of the $y=0$ plane, where density enhancement and velocity arrows are instead shown in the plane defined by the current sheet's lengthwise axis. The transparent yellow surface encloses the super-magnetosonic flow ($M_{A}>1$). We can see that the acceleration of plasma is predominantly confined along the current sheet length, with limited out of plane effects, which is consistent with the directionality of spine-fan reconnection shown in Figure \ref{fig:rcn}. 
} \label{fig:reversal_flow}
\end{figure*}

The spine-fan reconnection not only redistributes magnetic flux about the null point but also plasma, with the establishment of a system of flow through the current sheet (Figure \ref{fig:reversal_flow}).
It is not a simple \lq{disc-shaped}\rq{} outflow, relieving the compression of the implosion symmetrically, but rather plasma is directed predominantly along the current sheet axis as this is the dominant direction of the Lorentz force associated with sharply curved reconnected fieldlines. Along the outflow plasma is accelerated to super-magnetosonic speeds in two \textit{reconnection jets}. The jets are bounded by standing slow mode MHD shock fronts which separate the fieldlines in the inflow regions and the outflow regions and are analogous to those found in Petschek's 2D model \citep{1964NASSP..50..425P}.

As the outflow evolves, we observe the pooling of hot, over-dense plasma at the heads of the reconnection jets {leading to back-pressures that choke off the outflow. This is accompanied by} the relief of (plasma) compression in the current sheet itself via the expulsion of the excess plasma swept in by the initial implosion. A plasma rarefaction forms in the inflow lobes, with plasma accelerated through the current sheet unreplenished in the absence of continual driving flows. 
The consequence of this plasma and flux redistribution is that the balance of forces acting across the spine and fan planes  (the collapse of which is necessary to sustain the reconnection and current sheet) must change as reconnection proceeds. 

The competition of these effects is the source of dynamism in Oscillatory Reconnection. 
Shortly after the initial collapse halts, strong plasma pressure gradients build up at the jet heads exerting a force against the outflow. In conjunction with deleterious effects in the inflow region (such as the diffusion/redistribution of the driving magnetic field and current sheet decompression) this leads to a deceleration of the jets, which leads in turn to the formation of fast MHD shocks (termination shocks). These refract magnetic field towards the shock's tangent with increases in overall plasma pressure and field strength. This bending of the field away from the normal gives rise to the opposite polarity deflection currents which were earlier identified in Figure \ref{fig:current2}. These fast, nonlinear waves formed in the jet head travel in toward the null, and
act implosively with similar physics to the initial collapse. Thus, a nonlinear fast wave collapses in a quasi-1D fashion, altering the magnetic field in the vicinity of the null, first undoing the spine-fan collapse of the initial implosion, then overshooting into a secondary collapse which pushes the spine and fan together in the opposite sense. This reverses the sign of current at the null, and thus reverses the sense of the reconnection. With the completion of the reversal and the secondary implosion, we see the establishment of new (weaker) reconnection flows which drive the next reversal via the same processes.

\subsection*{Wave generation}
\begin{figure*}
    \centering
    \includegraphics[width=0.875\linewidth]{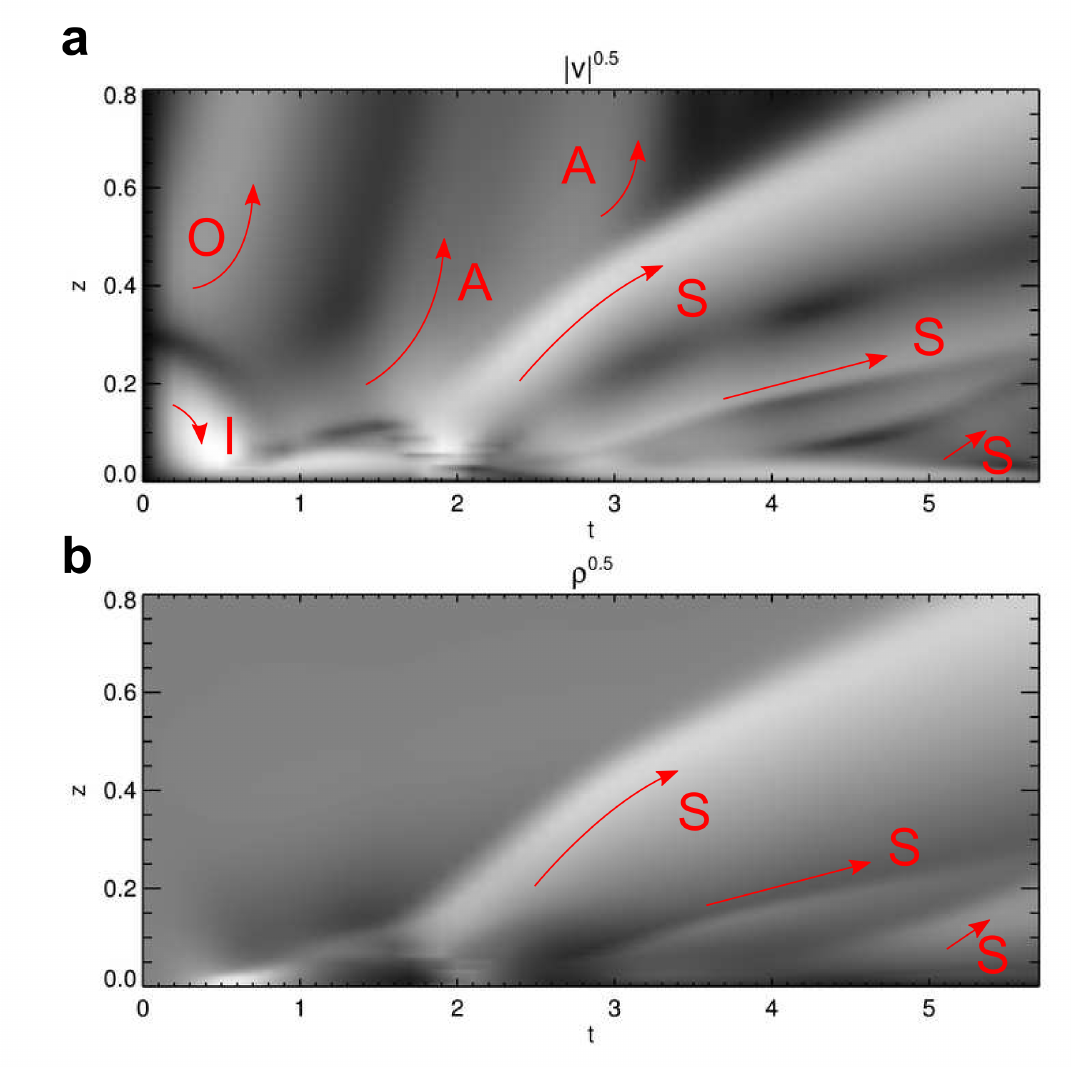} 
\caption{Time distance diagrams showing the evolution of velocity magnitude $|v|$ \textbf{(a)} and density $\rho$  \textbf{(b)} along the $z$-axis (the approximate position of the spine away from the collapse). The intensity {of both quantities} is scaled as the square-root to enhance contrast between different features. 
} \label{fig:td}
\end{figure*}
\begin{figure*}
    \centering
    \includegraphics[width=0.875\linewidth]{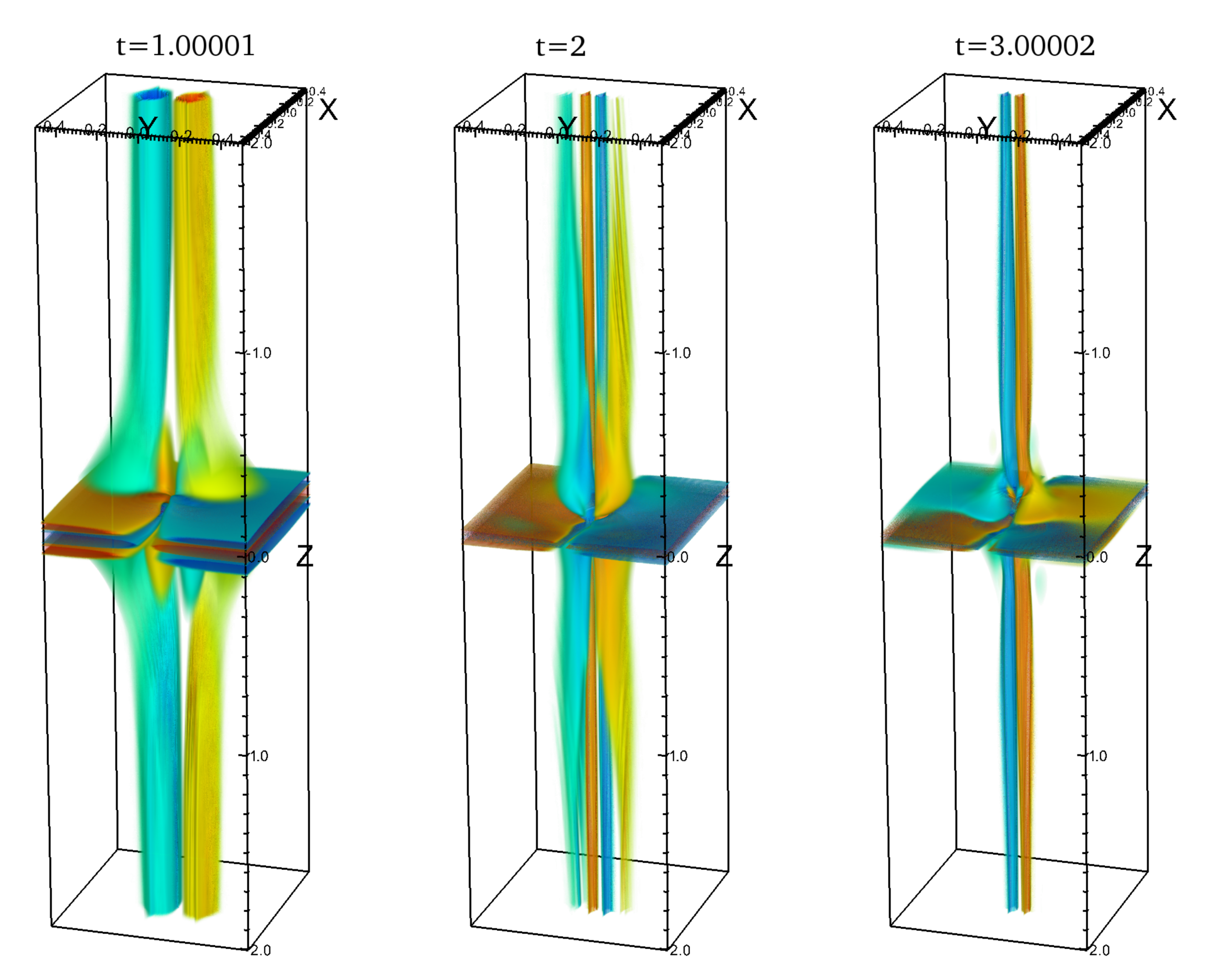} 
\caption{ Vorticity tube formation about the spine and in the fan plane over time ($\mathbf{\omega}\cdot\hat{\mathbf{B}}$) (see also the supplementary movie). The vorticity is associated with propagating Alfv\'en waves. 
} \label{fig:vort}
\end{figure*}
\begin{figure*}
    \centering
    \includegraphics[width=0.875\linewidth]{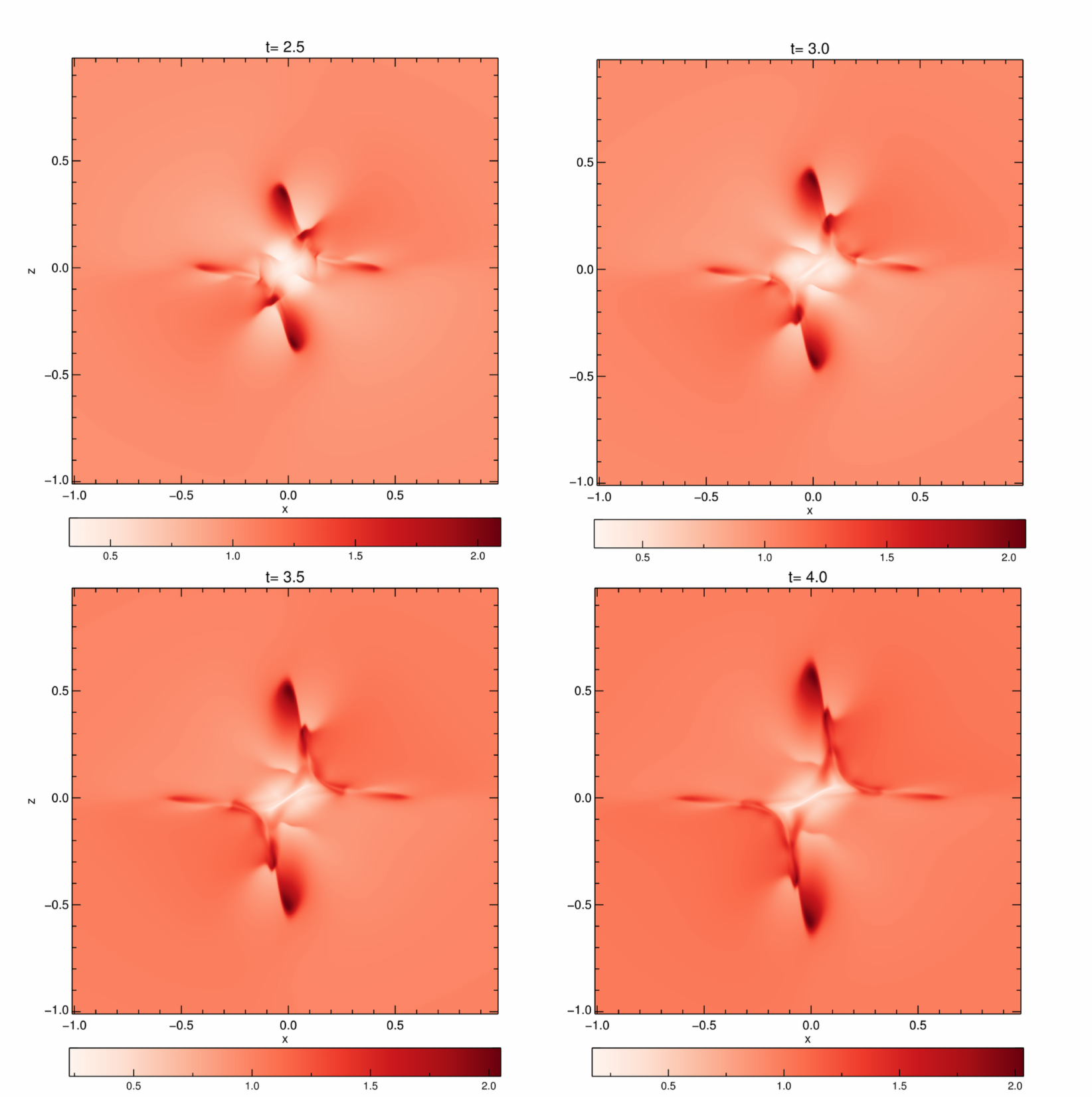} 
\caption{Propagating density fronts in the $y=0$ plane at selected times corresponding, which correspond to slow magnetoacoustic pulses. See also the corresponding movie.
} \label{fig:densfronts}
\end{figure*}

In these simulations the reconnection only occurs in a small region near the null point. However, the global effects of the magnetic field restructuring are not artificially confined to this area by computational boundaries,
but rather may escape the vicinity of the current layer as freely propagating waves.
We find that many escaping MHD waves are generated and highlight a selection here.

Figure \ref{fig:td}a,b shows time-distance diagrams which trace disturbances seen in $|{\bf v}|$ and $\rho$ propagating along the positive $z$-axis.  A number of different wavefronts are visible. The features labelled \textit{I} and \textit{O} respectively correspond to the incoming and outgoing waves generated by our initial perturbation, i.e., they are not generated by reconnection.

We identify the features marked \textit{A} as Alfv\'en waves. They propagate at the linearly increasing background Alfv\'en speed, which leads to the exponential profile along its fronts (by the solution of $dz/dt=|B_{z}|=2z$). They are, as characteristic of Alfv\'en waves, incompressible and are as such not visible in the figure for $\rho$ (Figure \ref{fig:td}b). These waves are also manifest in the generation of oppositely signed vorticity tubes in the reconnection region which propagate up along either side of the spine line (Figure \ref{fig:vort}). Vorticity 
propagation is a marker of Alfv\'en waves in 3D MHD, as it corresponds to a rotational disturbances of the magnetic field which allows magnetic tension to act as the predominant restoring force giving rise to the key properties of the wave \citep{2011SSRv..158..289G,2012A&A...545A...9T,2013ApJ...776L...4S}. These waves are   generated by the swaying of the spine line and  neighbouring fieldlines from side to side along the $x$-axis during each reversal event, which produces two  counter rotational vortices in the flow pattern surrounding the spine (somewhat reminiscent of the \lq{$m=1$}\rq{} kink wave in coronal loops). In this way the vorticity produced by the magnetoacoustic modes driving the swaying of the spine to the Alfv\'en mode in a process of \textit{mode conversion}.

The features marked $S$ are identified as {large-amplitude, anharmonic} slow-mode waves {--} due to their magnetoacoustic propagation speed (always lower than the Alfv\'en speed) {--} and have signal present in both the velocity and density diagrams. They are excited in the outflow regions of the reconnection jets (i.e., both the slow waves and termination shock have the same source). Owing to their strong nonlinearity {these anharmonic}  waves not only propagate plasma compression but actually transport material from the reconnection outflow away from the null and out along the spine axis (and fan plane). 
The  profile of material carried away from the null in $y=0$ at $t=4$ by such waves is also shown in Figure \ref{fig:densfronts}; note the multiple, large amplitude fronts along the spine ($x\approx0$) and fan ($z\approx0$) ({which indicates these waves propagate in a field-aligned sense, as expected of slow waves - see also the corresponding movie}), and the under-dense region that remains in the vicinity of the null point {as a consequence of the material transport}.

\section{Discussion}\label{discussion}

We have demonstrated unambiguously that magnetic reconnection at a 3D null point, driven by a perturbation of finite duration, naturally proceeds in a time-dependent, oscillatory manner. We find that the process consists of four key elements, namely (i) the creation of current sheets by MHD waves in a process of 3D null collapse, (ii) flux and plasma redistribution via localised spine-fan reconnection, (iii) associated reconnection reversals due to back-pressure-driven overshoots, and (iv) energy escape processes in the form of wave generation. 

The identification of 3D Oscillatory Reconnection is a milestone in our understanding of energy release in high magnetic Reynolds number plasmas,
demonstrating that reconnection triggered in an \emph{aperiodic} manner  both produces \emph{periodic}  reconnection reversals and \emph{periodically} excited propagating MHD waves  due to a self-generated oscillation.
Both the inherent periodicity of 3D reconnection events and the generation of waves may {play a role in explaining} the observed quasi-periodicity in solar and stellar flare emission \citep{2009SSRv..149..119N,2016SSRv..200...75N,2016SoPh..tmp..147V}. 
Understanding the physical mechanism underpinning this quasi-periodicity is crucial for developing the next generation of flare models
and for exploiting these observations as a diagnostic tool (via magneto-seismology, see \citealt{2012RSPTA.370.3193D}). 
{At present, there are a number of competing physical mechanisms that may explain Quasi-Periodic Pulsations (QPPs) including Oscillatory Reconnection, the magnetic tuning fork \citep{2016ApJ...823..150T}, self-oscillatory process  such as \textit{magnetic-dripping} \citep{2010PPCF...52l4009N}, and dispersive wave trains \citep{2004MNRAS.349..705N} (see also the aforementioned review papers). Our work finds that Oscillatory Reconnection may proceed  at fully 3D nulls and therefore is one potential mechanism that may provide an explanation for QPPs{, since a subset of  flaring events involve truly 3D null point geometries as opposed to quasi-2D X-lines (such as seen in e.g. \citealt{2009A&A...493..629Z,masson2009,2012ApJ...746...19Z}).}}

The initial conditions presented in this paper {were chosen by experimentation to} correspond physically to an MHD pulse impinging on the null which is sufficiently energetic (relative to the plasma pressure and resistivity) that it implodes in a nonlinear fashion as per \citet{1996ApJ...466..487M} and forms a current sheet {which then goes on to participate in Oscillatory Reconnection}. Pulses that are not sufficiently energetic, e.g. lower amplitudes (for the same pressure and resistivity), remain linear and do not perturb the spine and fan asymmetrically. They therefore do not form current sheets but rather ring currents, which evolve in a qualitatively different fashion to the system presented here. As our experiment is free of global scales its applicability to a given scenario, {such as determining its period for a typical solar event}, must be addressed by combining situation specific, global modelling with analysis of observational and experimental data. {We expect that}, in the solar atmosphere, given the abundance of energetic waves \citep[e.g.][]{2017SoPh..292....7L}, flux emergence events \citep[e.g.][]{2013ApJ...778...99L,2014ApJ...787...46L}, low pressures and low resistivity, the nonlinear regime of null point collapse is expected to be accessible {and so Oscillatory Reconnection should proceed in a qualitative sense as presented here}. {Determining the specific periods accessible by the mechanism in a solar scenario will require such models which place observational restrictions on permissible null geometries, driving energies, and plasma pressures, whilst also accounting for gravitational stratification.}

Furthermore, MHD waves are observed to be ubiquitous in the solar atmosphere \citep{2007Sci...317.1192T,2011Natur.475..477M,2015NatCo...6E7813M}. 
Understanding these waves is essential since they are believed to play a central role in the atmospheric mass and energy balance; however the nature of their genesis is an outstanding question.
The generation of mass-transporting slow modes due to reconnection may explain the origin of \textit{propagating intensity disturbances} observed in the open-field corona \citep{1998ApJ...501L.217D,2011SSRv..158..267B}.
Additionally, models predict that the dissipation of torsional Alfv\'en waves could provide significant chromospheric and coronal heating \citep{2010ApJ...712..494A}.
Thus, the natural excitation of torsional Alfv\'en waves by Oscillatory Reconnection
is especially interesting given that twist and torsional motions have recently been shown to be highly-prevalent in the low solar atmosphere \citep{2014Sci...346D.315D}.

There are also significant implications for the restructuring of the global field in response to the reconnection process. For example, reconnection at a 3D null point in the Sun's atmosphere can facilitate the exchange of plasma between magnetic fieldlines that are closed (connect at both ends to the solar surface) and those that are open to interplanetary space \citep{pontin2013}, driving material outflows \citep{bradshaw2011}. Our results thus provide a potential mechanism for (quasi-)periodic ejections observed in solar jets \citep{2012A&A...542A..70M,2014A&A...562A..98C,2014A&A...567A..11Z,2016SoPh..291..859Z}.

{Additionally},
{to our knowledge} Figure \ref{fig:reversal_flow}  is the first detailed consideration of {Petschek-like} reconnection jets in a {spine-fan} system, replete with standing slow shocks and fast termination shocks, with particular implications for non-thermal particle acceleration. 
{We note that the \textit{steady-state} Petschek reconnection in uniform-resistivity MHD is widely considered unattainable, with the requirement of an anomalous resistivity, localised at the null, to act as an obstacle in the flow (e.g., \citealt{2001PhPl....8.4729B}). Here we stress that the \textit{finite-duration} \lq{Petschek-like}\rq{} system obtained in this work, with uniform resistivity,   is intimately connected to the time-dependent nature of our problem.
}

{{Finally, we note that previous studies involving (2D) null collapse were chiefly motivated in terms of examining the possibility of the collapse to small scales and large current densities yielding favourable scaling laws of the reconnection rate with resistivity. Early results in the zero-$\beta$ limit suggested a `fast' scaling, with  decay rates of the perturbations scaling as $\sim|\ln\eta|^2$ in the linear regime and becoming virtually independent of $\eta$  in the nonlinear case (so-called \textit{super-fast} reconnection as per \citealt{1982JPlPh..27..491F,1996ApJ...466..487M}). However, studies at finite-$\beta$ suggest that in the corona back-pressure in the current layer may halt the collapse before it enters a regime of fast reconnection. There are a number of possible resolutions - in MHD this includes the possibility of strong nonlinear effects associated with highly energetic perturbations mitigating the effects of pressure leading to secondary, faster phases of reconnection, as hypothesised by \citet{1996ApJ...466..487M} and discussed in \citet{priestandforbes}. There also exists the possibility that even if the collapse is pressure-limited, provided the sheet width can reach kinetic length scales, then the resulting microphysical reconnection rates will be sufficiently fast. In the context of 2D null point collapse, the fast rates of collisionless reconnection  have been demonstrated by particle-in-cell simulations in a series of papers by Tsiklauri and collaborators \citep{2007PhPl...14k2905T,2008PhPl...15j2902T,2008PhPl...15k2903T,2014PhPl...21a2901G,2016A&A...595A..84G}. As such, an important outstanding issue is the efficiency of 3D null collapse as a reconnection and dissipation mechanism. In this paper, 3D null collapse features as the means of establishing a localised current sheet which precipitates Oscillatory Reconnection, which itself is our main focus. We do however note that we have found that three-dimensional null collapse is at least qualitatively similar to the known 2D behaviour. A detailed, quantitative study of the scaling of the 3D null collapse mechanism, and it's relation to the 2D case, will be important to undertake in future.}}

\section*{Acknowledgements}
The authors acknowledges generous support from the  Leverhulme Trust and this work was funded by a Leverhulme Trust Research Project Grant: RPG-2015-075. The authors acknowledge IDL support provided by STFC. The computational work for this paper was carried out on HPC facilities provided by the Faculty of Engineering and Environment, Northumbria University, UK.

\section*{Author Information}
The authors declare no competing financial interests or other conflicts of interest.
Correspondence should be addressed to jonathan.thurgood@northumbria.ac.uk

\bibliography{references}

\appendix
\section{Solver - Lare3d code}
The simulation is the numerical solution of the nondimensional, resistive MHD equations:
\begin{eqnarray}
\frac{\mathrm{D}\rho}{\mathrm{D}t}&=&-\rho \nabla\cdot \mathbf{v}\\
\frac{\mathrm{D}\mathbf{v}}{Dt}&=&\frac{1}{\rho}(\nabla\times\mathbf{B})\times\mathbf{B}
-\frac{1}{\rho}\nabla p + \mathbf{F}_{shock}\\
\frac{\mathrm{D}\mathbf{B}}{\mathrm{D}t}&=&(\mathbf{B}\cdot\nabla)\mathbf{v}-\mathbf{B}
(\nabla\cdot\mathbf{v})-\nabla\times(\eta\nabla\times\mathbf{B})\\
\frac{\mathrm{D}\epsilon}{\mathrm{D}t}&=&-\frac{p}{\rho}\nabla\cdot\mathbf{v}+\frac
{\eta}{\rho}j^{2} + \frac{\mathbf{H}_{visc}}{\rho}\\
\mathbf{j} &=& \mathbf{\nabla}\times\mathbf{B}\\
\mathbf{E} &=& -\mathbf{v}\times\mathbf{B}+\eta\mathbf{j}\\
p &=& \epsilon\rho\left(\gamma-1\right)
\end{eqnarray}
which are solved using the \textit{Lare3d} code. All results presented are in non-dimensional units. Algorithmically, the code solves the ideal MHD equations explicitly using the Lagrangian remap approach and includes the resistive terms using explicit subcycling \citep{2001JCoPh.171..151A,2016ApJ...817...94A}. The solution is fully nonlinear and captures shocks via an edge-centred artificial viscosity approach \citep{1998JCoPh.144...70C}, where shock viscosity is applied to the momentum equation through $\mathbf{F}_{shock}$ and heats the system through $\mathbf{H}_{visc}$. Extended MHD options available within the code, such as the inclusion of Hall terms, were not used in these experiments. Full details of the code can be found in the original paper \citep{2001JCoPh.171..151A} and the users manual. Requests for access to the code should be addressed to its developers.

\section{Initial Conditions}

A summary of the (nondimensional) initial conditions on the primitive variables, discussed in the main document, is as follows:
\begin{eqnarray}
p &=& 0.005 \\
\rho &=& 1 \\
\mathbf{v} &=& \mathbf{0}\\
\mathbf{B} &=& \mathbf{B}_{0} + \mathbf{\nabla}\times\mathbf{A}_{1} \\
\mathbf{B}_{0} &=& \left[x,y,-2z \right] \\
\mathbf{A}_{1} &=& \psi\,\mathrm{exp}\left[\frac{-\left(x^2+y^2+z^2\right)}{2\sigma^2}\right] \hat{\mathbf{y}} \\
\psi &=& 0.05\\
\sigma &=& 0.21
\end{eqnarray}
with resistivity taken uniformly as $\eta=10^{-3}$ throughout.  The ratio of specific heats is taken as $\gamma=5/3$. 

\section{Boundary conditions and reflectivity testing}

The boundary conditions are taken as no-slip ($\mathbf{v}=\mathbf{0}$) with zero-gradient conditions on $\mathbf{B}$, $\rho$ and $p$. A simulation was run with no perturbation to check that the boundaries do not launch spurious waves into the domain, and to check the overall stability of the setup. 

Special care must be taken so that outgoing waves, whether from the initial perturbation or those subsequently generated, do not reflect off the domain boundaries and return inwards to influence our solution at the null. During our initial testing we found that open boundaries and artificial damping zones are imperfect strategies for removing outgoing waves in linear null point geometries. As such the only robust method of avoiding reflections influencing the solution is to simply place the boundaries sufficiently far away from the null point that the  shortest possible signal travel time from the initialisation site to the boundary and back is in excess of the period we wish to study. Computationally, this is not a trivial task.
The shortest signal travel time is governed by the Alfv\'en speed which increases linearly with distance from the null point (at the center of the domain). Thus, the overall travel time to the boundary increases only logarithmically with increasing distances to the boundaries. Boundaries must be placed at vast distances whilst maintaining sufficient resolution in the dynamic and diffusive regions close to the null point. 

For the results shown here, we placed the $x$,$y$ and $z$ boundaries at $\pm40$. The shortest possible signal travel time permitted is that of a wave propagating at the Alfv\'en speed directly along the spine (where $B_z$ increases as $2z$), and so by separation of variables the travel time from the outer edges of the initial perturbation $z\approx\pm0.25$ out to the boundary is $t_{out}\approx{0.5}\ln(40/0.25)=2.53$. The fastest possible reflection will return to the diffusion dominated region about the null (estimated at the length of the initial current sheet, $\approx0.18$) after a further $t_{in}\approx{0.5}\ln(40/0.18)=2.70$, yielding the shortest possible theoretical reflection time of $t_{\mathrm{reflect}}\approx5.23$. 
This was also practically tested with a \lq{boundary convergence test}\rq{} where simulations were run with boundaries which were closer to the null (hence having shorter reflection times). We found that results only differed at times in excess of the theoretical reflection time calculated as above for each case. In these test cases it was often possible to track reflections along paths using time distance diagrams as per Figure \ref{fig:td}a, which further confirmed these theoretical times. The maximum boundary length used ($\pm40$) was a compromise between maximising the reflection time and maintaining the resolution available close to the null -- note that  doubling the distance to the boundary will only increase the time by $\ln(2)$.
Thus, by simple wave theory, the fastest possible theoretical reflection time is  $t_{\mathrm{reflect}}\approx5.23$ and so the non-interference of reflections during the initial implosion, the first reversal, and the initial stages of the second reversal is guaranteed. Therefore, the self-generated oscillations of the null point are indeed due to reconnection itself. Further, we note in the later stages of the second reconnection reversal ($t>t_{\mathrm{reflect}}$) up to the end time of $t=6$ that no incoming fronts of significant intensity are detected in Figure \ref{fig:td}a, which is in fact taken along this shortest possible signal travel path.

\section{Grid geometry and convergence testing}

To accommodate the vast distances to the boundary whilst sufficiently resolving the diffusion at the null point and the wave dynamics in the surrounding region a grid stretching scheme is employed. 

The grid is stretched according to a variation on the scheme of \cite{1971LNP.....8..171R}, namely the cell boundary positions $x_{b}$ along the $x$-direction  are distributed according the transformation:
\begin{eqnarray}
x_{b} &=& 40 \left\lbrace 1 + \frac{\sinh\left[\lambda\left(\xi_{i}-\Gamma\right)\right]}{\sinh\left[\lambda\Gamma\right]} \right\rbrace - 40 \\
\Gamma &=& \frac{1}{2\lambda}\ln\left[\frac{1+\left(e^{\lambda}-1\right)0.5}{1-\left(1-e^{-\lambda}\right)0.5}\right]
\end{eqnarray}
where $\xi_{i}$ is a uniformly distributed computational coordinate $\xi\in[0,1]$ subdivided amongst the number of cells used in the $x$ direction. The degree of grid clustering at the origin is controlled by the stretching parameter $\lambda$. Likewise, the same form and parameters are used for the distribution of cells in $y$ and $z$. This choice thus clusters cells near the origin (null point).  

In our final simulations we take $\lambda=15.5$, and use $512^3$ computational cells ($512$ in each direction). This yields maximal resolution of $\Delta{x}\approx0.0011$  close to the null point itself (i.e., this simulation, with the far boundaries, would require approximately $(8\times10^{4})^{3}$ cells on a uniform grid). The sufficiency of the resolution in the regions surrounding the null which contribute to the solution was tested in two ways. First, we ran quantitative convergence tests on the initial implosive phase (up to approximately $t=0.6$) on a version of the simulation using uniform grids of successively finer resolutions (facilitated with much closer boundaries, as the reflection time is not prohibitive for such purposes). {Using this approach we were able to ensure that} the collapse is adequately resolved, and the solution converges. {These simulations were} also compared to the same stages of the final simulation using the stretched grid (showing good agreement). The resolution close to the null gives approximately $50$ grid points across the width of the current sheet (diffusion region) itself, ensuring the current is not concentrated at the grid scale, meaning that unphysical numerical diffusion is negligible.  Our second approach was to perform lower resolution simulations using the grid stretching ($128^3$ and $256^3$ cells), which produced similar results. We note that below $256^3$ cells the solution in the wave dominated region surrounding the null begins to deteriorate (e.g., the structure across the jets). Additionally, the ability of the stretched geometry to maintain the self-similar nature of implosions and explosions was tested by comparison of the stretched grid to uniform grid problems in runs with Sedov and MHD blasts, and with the successful reconstruction of the implosion/explosion of cylindrical MHD fast waves problem of \citep{2009A&A...493..227M}, a 2D problem which originally required a vast uniform grid, on a small stretched grid.

The final simulations required approximately 5 days of run time on $16$ nodes of hyperthreaded $28$ core $2.4\mathrm{GHz}$ Intel Xeon processors ($896$ threads), with high-speed omnipath interconnect between the nodes and $64\mathrm{Gb}$ RAM per node. The use of $512^3$ element arrays for a simulation of 8 primitive variables (plus arrays for ancillary calculations) does not present a significant RAM overhead for modern nodes, rather the large computation time is due to the requirement of a large number of simulation cycles due to the physical parameters and grid dictating very small hydrodynamic and resistive timesteps.

\section{Tracking connectivity change with tracer particles}\label{seedsapp}
Figure \ref{fig:rcn} was created as follows: (initially) magnetically connected fluid elements at either side of the flux tubes selected were tracked by writing a custom tracer module for Lare3d (i.e. fluid elements are followed in a Lagrangian sense as passive particles in the flow). This module is called after each hydrodynamic step of the main code and simply updates each particle position with an Eulurian push based on the hydrodynamic timestep and tri-linearly interpolated velocity at the particle's previous position ($\mathbf{x}_{\mathrm{t+1}}=\mathbf{x}_{\mathrm{t}}+{{\Delta}\mathrm{t}}\,\mathbf{v}$). 
The tracked particle positions are used then as instantaneous seed points for fieldline tracing using the inbuilt features of \textit{VisIt} \citep{HPV:VisIt} at each time frame shown in Figure \ref{fig:rcn} and its supplementary movie. After the bifurcation, new seeds are traced from either end of the resultant four flux ropes which are corresponding to fluid elements which are initially connected at $t=2.5$. These seeds are used for tracing in the same fashion for $t>2.5$. The particle module was tested exactly in linear flow fields. The tracers were also found to behave in the expected fashion for a number of test problems including MHD wave propagation, the Kelvin-Helmholtz instability, and blast waves.


\end{document}